\documentclass[amsmath,amssymb,twocolumn,amsfonts,twocolumn]{revtex4}
\usepackage{graphicx}
\def \beq {\begin{equation}}
\def \eeq {\end{equation}}
\def \tr {\rm Tr}

\begin{document}
\title{Radical-ion-pair reactions are the biochemical equivalent of the optical double slit experiment}
\author{Iannis K. Kominis}

\affiliation{Department of Physics, University of Crete, Heraklion
71103, Greece}

\begin{abstract}
Radical-ion-pair reactions were recently shown to represent a rich biophysical laboratory for the application of quantum measurement theory methods and concepts. We here show that radical-ion-pair reactions essentially form a non-linear biochemical double slit interferometer. Quantum coherence effects are visible when "which-path" information is limited, and the incoherent limit is approached when measurement-induced decoherence sets in. Based on this analogy with the optical double slit experiment we derive and elaborate on the fundamental master equation of spin-selective radical-ion-pair reactions that covers the continuous range from complete incoherence to maximum singlet-triplet coherence.
\end{abstract}

\maketitle
\section{Introduction}
The double slit experiment is the archetypal system manifesting the core concepts of quantum physics. We here show that Nature has already designed biologically significant chemical reactions that act as a double slit interferometer.
Spin-selective radical-ion-pair reactions are perhaps the only example in chemistry where
spin degrees of freedom and their relatively small interaction energy can have a
disproportionately large effect on the outcome of chemical reactions. Their study is at the core of spin chemistry \cite{steiner}, by now a mature research field directly related to photochemistry \cite{turro} and photosynthesis \cite{blankenship}. Radical-ion-pair reactions determine the late-stage dynamics in photosynthetic reaction centers \cite{boxer1,boxer2}, and are understood to underlie the avian compass mechanism, i.e. the biochemical compass used by migratory birds to navigate through the geomagnetic field \cite{schulten1,ww1,ritz,ww2,schulten2,maeda,rodgers}. A deeper understanding of the quantum dynamics inherent in radical-ion-pair reactions will thus enhance or even fundamentally alter our understanding of these biological processes. Other biochemical systems studied from the quantum interference perspective include electron transfer pathways \cite{skourtis}, and in a much broader context  the coherent control of molecular dynamics \cite{shapiro} and the femtosecond control of chemical reactions \cite{gerber,stavros}.

Radical-ion pairs are biomolecular ions created by a charge transfer from a
photo-excited D$^*$A donor-acceptor molecular dyad DA, schematically described by the reaction ${\rm DA}\rightarrow {\rm D^{*}A}\rightarrow {\rm D}^{\bullet +}{\rm A}^{\bullet -}$, where the two dots represent the two unpaired electrons. The magnetic nuclei of the donor and acceptor molecules couple to the two electrons via the hyperfine interaction, leading to singlet-triplet mixing, i.e. a coherent oscillation of the spin state of the electrons. Charge recombination terminates the reaction and leads to the formation of the neutral reaction products. It is angular momentum conservation at this step that empowers the molecule's spin degrees of freedom to determine the reaction's fate: only singlet state radical-ion pairs can recombine to reform the neutral DA molecules, whereas triplet radical-ion pairs recombine to a different metastable triplet neutral product.

Theoretically, the fate of radical-ion-pair reactions is accounted for by the time
evolution of $\rho$, the density matrix describing the spin state of the molecule's two electrons and any number of magnetic nuclei. It was recently shown \cite{kominis_PRE} that quantum measurement dynamics are central in these biochemical reactions. A new master equation was derived \cite{kominis_PRE} based on quantum measurement theory, as the radical-ion-pair recombination process was interpreted to be a continuous quantum measurement of the spin state of the pair's electrons. This master equation accounts for the spin decoherence of unrecombined radical-ion pairs. The kinetics of the recombination process, i.e. the loss of radical-ion pairs due to the formation of neutral products, must also be taken into account. The treatment of this problem in \cite{kominis_PRE} applies only to the case of maximal singlet-triplet coherence. Describing the continuous range of partial coherence down to the other extreme of maximal incoherence is what we need in order to provide a complete description of this system. We will here develop exactly this complete theoretical description of radical-ion-pair reactions, shown to be an almost perfect analogy of Young's double slit experiment with partial "which-path" information. In Section II we recapitulate the open-quantum-system approach to radical-ion-pair reactions, elucidating the physics of the unavoidably present decoherence due to the intra-molecule measurement dynamics introduced in \cite{kominis_PRE} and subsequently studied also in \cite{JH}. In Section III we introduce the reaction terms in the two extremes of maximum coherence and maximum incoherence. We then study the general case of partial coherence that naturally leads to the full master equation presented in Section IV along with some of its predictions in very simple cases. Finally in Section V we elaborate in detail on several aspects of the analogy with the photon's double-slit experiment.
\section{Radical-ion pairs as an open quantum system}
\begin{figure}
\includegraphics[width=8.5 cm]{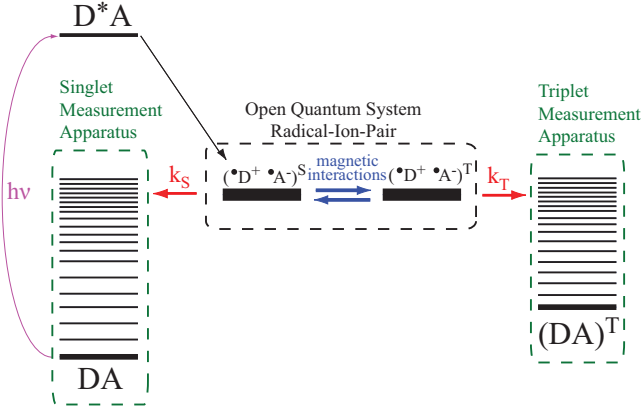}
\caption{(Color online) (a) Energy level structure of radical-ion-pair reaction dynamics. A photon excites the singlet neutral precursor molecule DA into D$^*$A, and a charge transfer creates the radical-ion pair.
The excited vibrational levels (DA)$^*$ of the neutral DA molecule form the measurement reservoir, which (i) acts as a measurement device for the radical pair's spin state, and (ii) acts as a sink of radical-ion pairs, i.e. in the event of recombination, the electron tunnels into a reservoir state and a fast spontaneous decay results in the ground state DA (which is the singlet product) and a photon emission. Similar for the triplet reservoir.}
\label{fig1}
\end{figure}

The reaction dynamics and energy levels of radical-ion pairs are depicted in Fig.\ref{fig1}.
We neglect effects like diffusion, collisions, spin relaxation etc and we only consider the fundamental quantum dynamics inherent in the recombination process. In all experimental studies we have a macroscopic number of radical-ion pairs. Each one of them is a single open quantum system regarding the spin degrees of freedom, because the recombination dynamics {\it inherent in each molecule} disturbs a what would be a unitary spin evolution. Moreover, due to recombination, radical-ion pairs form neutral chemical products, i.e. we deal with an ensemble with a changing number of open quantum systems, with these two aspects being intertwined. By properly identifying the system and the reservoir degrees of freedom in radical-ion pairs, we recently derived \cite{kominis_PRE} a {\it trace-preserving} master equation describing the evolution of the radical-ion-pair's spin state {\it until it recombines}:
\beq
{{d\rho_{\rm nr}}\over {dt}}=-i[{\cal H},\rho]-{{k_{S}+k_{T}}\over 2}\big(\rho Q_{S}+Q_{S}\rho-2Q_{S}\rho Q_{S}\big)\label{komME}
\eeq
Here ${\cal H}$ is the Hamiltonian embodying the magnetic interactions within the molecule, $Q_{S}$ is the singlet projection operator (the triplet projection operator is denoted by $Q_{T}$), and $k_S$ and $k_T$ are the singlet and triplet recombination rates, respectively. Equation \eqref{komME} embodies the {\it decoherence} or dephasing of the singlet-triplet coherence brought about by the intra-molecule continuous measurement induced by the recombination process. Due to this internal decoherence process, coherent superpositions are gradually turned into incoherent mixtures. To explain this in more detail, we first note that the singlet and the triplet reservoir enter the dynamics on an equal footing \cite{kominis_PRE}, the corresponding Lindblad terms being $-k_{S}(\rho Q_{S}+Q_{S}\rho-2Q_{S}\rho Q_{S})/2$ and $-k_{T}(\rho Q_{T}+Q_{T}\rho-2Q_{T}\rho Q_{T})/2$. Due to the completeness relation $Q_{S}+Q_{T}=1$, the sum of these terms results in the second, dissipative term of \eqref{komME}. Hence both reservoirs essentially measure one and the same observable, $Q_{S}$, the measurement result being $Q_{S}=0$ or $Q_{S}=1$. The total measurement rate is $(k_{S}+k_{T})/2$.
\begin{figure}
\includegraphics[width=8.5 cm]{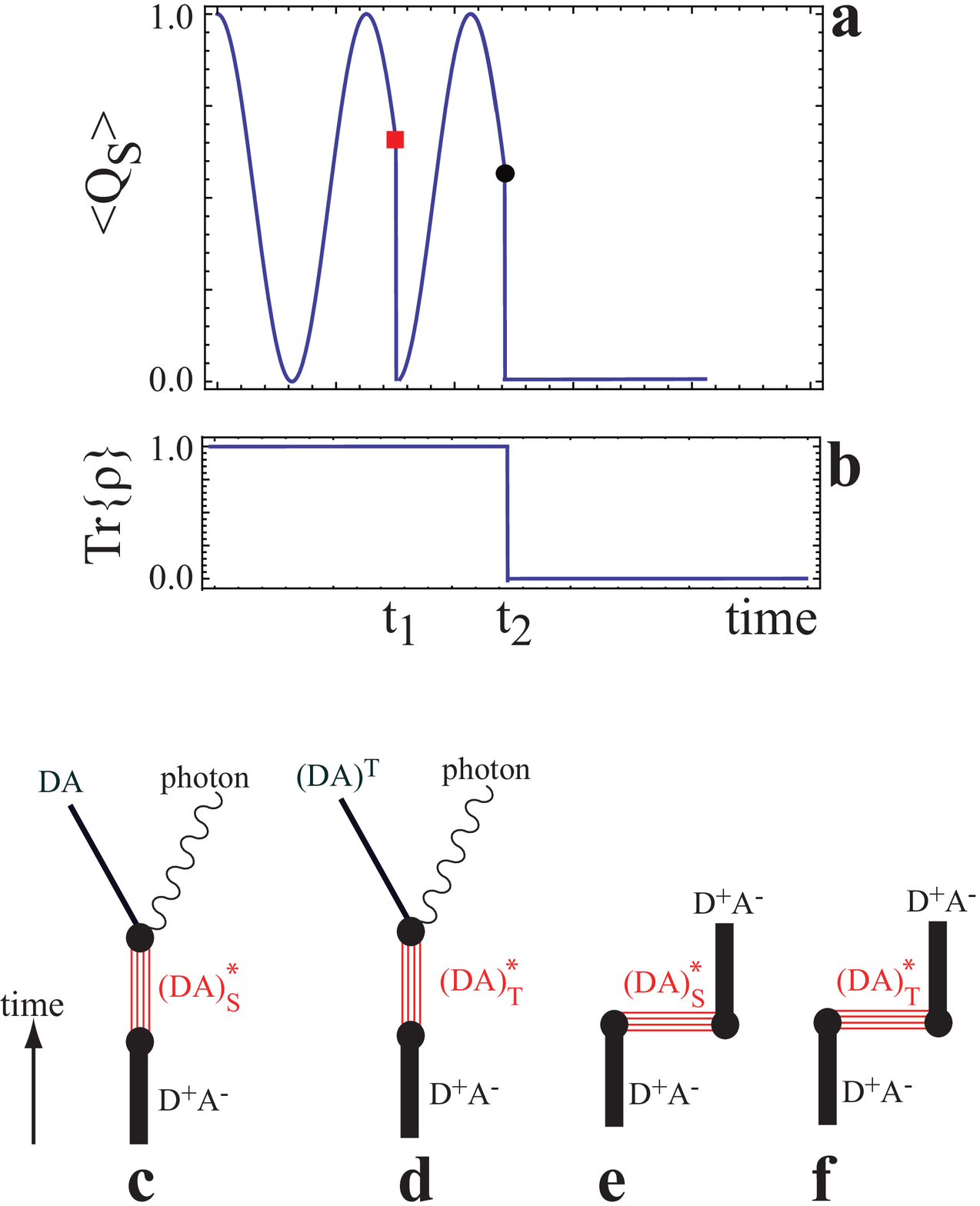}
\caption{(Color online) (a) Single-molecule quantum trajectory illustrating quantum jumps interrupting a unitary state evolution driven by the magnetic Hamiltonian ${\cal H}$. At some random time $t_1$ the measurement outcome is $Q_{S}=0$, and the radical-pair's spin state is projected (red square) to the triplet state, from which the unitary evolution commences. At some later time $t_{2}$ (black circle) the radical-ion pair charge recombines and disappears. (b) The recombination event changes the density matrix normalization from one to zero. (c) and (d) Feynman diagrams describing a singlet and a triplet recombination. (e) and (f) Feynman diagrams describing virtual tunneling to singlet or triplet reservoir states responsible for singlet-triplet decoherence.}
\label{feynman}
\end{figure}
This continuous measurement interrupts a what would be a unitary spin evolution driven by the magnetic Hamiltonian ${\cal H}$. In other words, this unitary evolution is interrupted by quantum jumps produced by the projectors $Q_{S}$ and $Q_{T}$. One illustrative example of such a single-molecule quantum trajectory is shown in Fig.2a. At some random time the radical-pair's state jumps to the triplet state ($Q_{S}=0$), and at some later time the radical-pair recombines, hence from that point on $\langle Q_{S}\rangle=\langle Q_{T}\rangle=\tr\{\rho\}=0$ (Fig.2b). The average of all such trajectories (without the recombination event) is embodied in the master equation \eqref{komME}. The recombination event is described by Feynman diagrams shown in Figs. 2c and 2d. The radical-ion pair state is coupled (by tunneling) to the reservoir state, which then emits a photon (or phonon) and the neutral product (ground state of DA or (DA)$^T$). The dephasing produced by the quantum jumps is described by a diagram like the one in Figs. 2e and 2f. Here the radical-pair "momentarily" tunnels to a singlet (triplet) reservoir state, from which it tunnels back to a singlet (triplet) radical-ion-pair state and the evolution commences on from there. This coupling to and back from the reservoir is taken into account by the second-order perturbation in the system-reservoir coupling that leads to \eqref{komME}, as shown in \cite{kominis_PRE}. To recapitulate, the excited vibrational levels of the singlet DA molecule and the triplet (DA)$^T$ molecule form the singlet and triplet reservoirs, respectively. Virtual transitions to both singlet and triplet reservoir states are responsible for singlet-triplet (S-T) decoherence, whereas real transitions to singlet (triplet) reservoir states cause singlet (triplet) recombination.

The state change of unrecombined radical-ion pairs due to the intra-molecule quantum measurement dynamics is one part of the whole picture. The other is the change of the ensemble density matrix due to radical-ion pairs leaving the ensemble, i.e. recombining (reacting) away. We now come to the description of the reaction terms.
\section{Reaction Terms}
Suppose that at time $t$ an {\it ensemble} of radical-ion pairs is described by $\rho_t$. Sure enough, in the following time interval $dt$ we will obtain $dn_{S}=k_{S}dt\tr\{\rho Q_{S}\}$ singlet products and $dn_{T}=k_{T}dt\tr\{\rho Q_{T}\}$ triplet products. In other words, at time $t+dt$ we will have $dn_{S}+dn_{T}$ less radical pairs. We emphasize that the measurement of these product molecules represents {\it classical information}, the acquisition of which or not {\it cannot have any back-action on the state of unrecombined radical pairs}, as much as the detection of photons on the observation screen beyond Young's double slit does not back-react on the quantum state of the rest of the photons flying through the double slit. The information on the product molecules  is in principle conveyed by the detection of $dn_{S}$ photons (or phonons) as depicted in Figs.\ref{feynman}c and \ref{feynman}d \cite{note0}. The question we will address is the following: how is $\rho_{t}$ consistently evolved into $\rho_{t+dt}=\rho_{t}+d\rho$? In order to simplify and motivate the present discussion we consider just two states \cite{note1}, the singlet $|S\rangle$ and the triplet $|T\rangle$, i.e. we consider a two-dimensional Hilbert space. Our results are generally valid for any radical-ion-pair Hilbert space.
\subsection{Maximum incoherence extreme}
Suppose we know that at time $t$ we have an incoherent singlet-triplet mixture, i.e. $\rho_{t}=\lambda_{S}|S\rangle\langle S|+\lambda_{T}|T\rangle\langle T|$, with $\lambda_{S}+\lambda_{T}=1$, or equivalently, using the projectors $Q_{S}=|S\rangle\langle S|$ and $Q_{T}=|T\rangle\langle T|$ we can write $\rho_{t}=\lambda_{S}Q_{S}+\lambda_{T}Q_{T}$. This is the maximum incoherence extreme, i.e. there is zero S-T coherence. We would then detect $dn_{S}=k_{S}\lambda_{S}dt$ singlet and $dn_{T}=k_{T}\lambda_{T}dt$ triplet products. Due to our knowledge of $\rho_{t}$, we would {\it also know for sure} that the detected $dn_S$ singlet and $dn_{T}$ triplet products {\it must have originated} from singlet state and triplet state radical-ion pairs, respectively. Thus we would write
\beq
d\rho_{\rm incoh}=-k_{S}dtQ_{S}\rho Q_{S}-k_{T}dtQ_{T}\rho Q_{T}
\eeq
That is, we project out the singlet and independently the triplet part of $\rho_{t}$ by the reacted fraction of singlet ($k_{S}dt$) and triplet ($k_{T}dt$) radical-ion pairs. Obviously $\tr\{d\rho_{\rm incoh}\}=-dn_{S}-dn_{T}$, as it should be.
\subsection{Maximum coherence extreme}
Suppose, on the other hand, that at time $t$ all radical-ion pairs are in the maximally coherent state $|\psi\rangle=(|S\rangle+|T\rangle)/\sqrt{2}$ \cite{note2}. Thus $\rho_{t}=|\psi\rangle\langle\psi |=(|S\rangle\langle S|+|T\rangle\langle T|+|S\rangle\langle T|+|T\rangle\langle S|)/2$. Since $\tr\{\rho_{t} Q_{S}\}=\tr\{\rho_{t} Q_{T}\}=1/2$, we expect $dn_{S}=k_{S}dt/2$ singlet and $dn_{T}=k_{T}dt/2$ triplet products. What information does $\rho_{t}$ convey about the possible precursors of the  $dn_S$ singlet and $dn_T$ triplet products? None. As the reacted radical-ion pairs cease to exist, so does the information about their particular quantum state just prior to recombination. So now, in order to update $\rho_{t}$ we have to remove the complete single-molecule density matrix $\rho_{t}/\tr\{\rho_{t}\}$ as many times as many products we measured, i.e. \beq
d\rho_{\rm coh}=-(dn_{S}+dn_{T})\rho_{t}/\tr\{\rho_{t}\}
\eeq
This is a crucial point. The spin state of the singlet or the triplet neutral product will indeed be $Q_{S}\rho_{t}Q_{S}/\tr\{\rho_{t}Q_{S}\}$ and  $Q_{T}\rho_{t}Q_{T}/\tr\{\rho_{t}Q_{T}\}$, respectively, but that does not imply that the state of the precursor radical-ion pair was a pure singlet or a triplet. It was a coherent superposition. In other words, coming back to the double slit analogy, observing a photon in the upper (lower) part of the observation screen does not imply that the photon had crossed the upper (lower) slit. That would obliterate the physical essence of quantum superpositions. Again, as in the previous case, $\tr\{d\rho_{\rm coh}\}=-dn_{S}-dn_{T}$, as it should be. It is finally noted that
the treatment of reaction kinetics in our first publication \cite{kominis_PRE} applies to just this special case of maximum singlet-triplet coherence.
\subsection{General case: partial coherence}
We now have to address the general case of partial coherence that spans the region between the above two extremes. Suppose that $\rho_t$ is an arbitrary density matrix in the radical-pair's spin space. What we need is a {\it measure} of singlet-triplet coherence of $\rho_{t}$, that will continuously span the intermediate region between zero coherence and maximum coherence. Before deriving this measure, we briefly mention the analogous situation in the optical double slit experiment. Photon interference with partially coherent light is now a textbook example \cite{knight}, the intensity at the observation screen is $I(\mathbf{r})=I_{1}+I_{2}+2\sqrt{I_{1}I_{2}}\gamma(x_{1},x_{2})$, where
$I_{j}\sim \langle |E(x_{j})|^{2}\rangle$ is the intensity at the observation point $\mathbf{r}$ that would result from only slit $j$ being open, with $j=1,2$, and $\gamma(x_{1},x_{2})$ the complex first-order coherence function, with $|\gamma|=1$ for complete coherence, $|\gamma|=0$ for complete incoherence and $0<|\gamma|<1$ for partial coherence.

The analogous measure of S-T coherence for radical-ion pairs is derived as follows. By multiplying $\rho_{t}$ from left and right by the unit operator, and replacing the latter by $Q_{S}+Q_{T}=1$, we can write
$\rho_{t}=\rho_{SS}+\rho_{TT}+\rho_{ST}+\rho_{TS}$, where $\rho_{SS}=Q_{S}\rho_{t} Q_{S}$, $\rho_{TT}=Q_{T}\rho_{t} Q_{T}$, $\rho_{ST}=Q_{S}\rho_{t} Q_{T}$ and $\rho_{TS}=Q_{T}\rho_{t} Q_{S}$. The coherence of $\rho_{t}$ is obviously $\rho_{ST}+\rho_{TS}$. Since $\rho_{ST}$ and $\rho_{TS}$ are 
traceless (because $Q_{S}Q_{T}=0$), we consider the trace of their product as the measure describing the "amount" of singlet-triplet coherence, i.e. the quantity $\tr\{\rho_{ST}\rho_{TS}\}$. We define the S-T coherence measure $p_{\rm coh}$ as
\beq
p_{\rm coh}={{\tr\{\rho_{ST}\rho_{TS}\}}\over {\tr\{\rho_{SS}\}\tr\{\rho_{TT}\}}}\label{pcoh}
\eeq
For the maximal coherence extreme it is $p_{\rm coh}=1$, while for completely incoherent mixtures we have $p_{\rm coh}=0$, with all other values covering the intermediate partial coherence regime. These properties of $p_{\rm coh}$ are proved in Appendix A.

So now, the only statement we can make about the change in the density matrix $d\rho$ consistent with the information at hand is that
\beq
d\rho_{r}=(1-p_{\rm coh})d\rho_{\rm incoh}+p_{\rm coh}d\rho_{\rm coh}
\eeq
This is the reaction term for a general state $\rho_{t}$ having partial S-T coherence. In other words, for partial S-T coherence, we only know the probability that, for example, a singlet product came from a singlet precursor radical-pair, or from an S-T coherent radical-pair. The former is $1-p_{\rm coh}$ and the latter is $p_{\rm coh}$.
\section{Master Equation}
To arrive at the complete theory, we have to add the change of the density matrix due to the recombined radical-pairs, $d\rho_{r}$, with the change of state of the unrecombined radical pairs, $d\rho_{nr}$, leading to the master equation
\begin{align}
{{d\rho}\over {dt}}&=-i[{\cal H},\rho]-{{k_{S}+k_{T}}\over 2}\big(\rho Q_{S}+Q_{S}\rho-2Q_{S}\rho Q_{S}\big)\nonumber\\&-(1-p_{\rm coh})(k_{S}Q_{S}\rho Q_{S}+k_{T}Q_{T}\rho Q_{T})\nonumber\\
&-p_{\rm coh}\big(k_{S}\tr\{Q_{S}\rho\}+k_{T}\tr\{Q_{T}\rho\}\big){\rho\over {\tr\{\rho\}}}\label{KOM}
\end{align}
This master equation embodies the fundamental quantum dynamics of spin-selective radical-ion-pair reactions.
Since $d\tr\{\rho_{\rm nr}\}=0$, it is seen that the trace of $\rho$ decays as it should, i.e. according to the number of recombined radical-ion pairs, $d\tr\{\rho\}=-dn_{S}-dn_{T}$. The singlet and triplet reaction yields are then
$Y_{S}=\int_{0}^{\infty}{dn_{S}}$ and $Y_{T}=\int_{0}^{\infty}{dn_{T}}$, respectively. The traditional theory \cite{trad} identically follows from \eqref{KOM} by setting $p_{\rm coh}=0$ for all times. That is, the traditional theory by design completely suppresses the effect of radical-pair spin coherence. We shall show (Section V.D.) that this is done by changing the state of unrecombined radical-ion pairs in an unphysical way.
\subsection{Predictions of Master Equation}
The predictions of this master equation and their comparison with the traditional paradigm is a rather long program to be undertaken elsewhere.  
We will here illustrate just one quite striking example on the differences between the new and the previous theoretical understanding  (a more recent approach by Jones \& Hore \cite{JH} offers similar predictions as the traditional theory). This example serves to unravel fundamental differences in the quantum mechanical understanding of radical-ion-pair reaction dynamics in a simple enough setting that will allow the analogy with the optical double-slit experiment to clearly illuminate these differences. 
\begin{figure}
\includegraphics[width=7.5 cm]{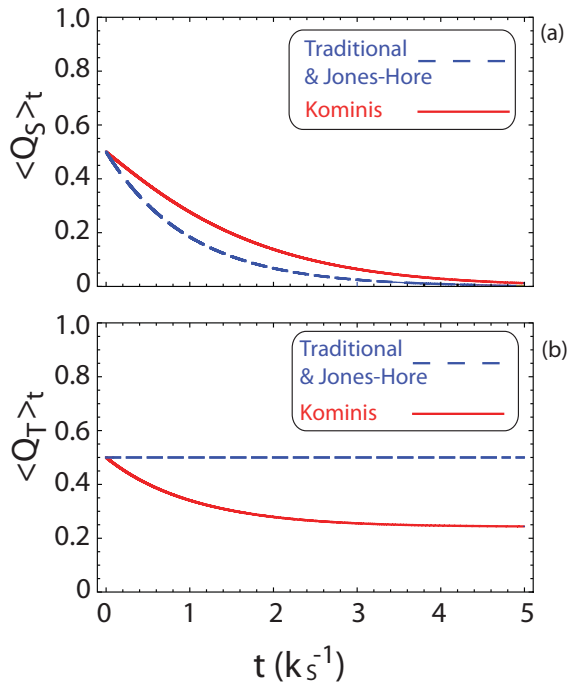}
\caption{(Color online) Radical-ion-pair recombination dynamics in the absence of singlet-triplet mixing. (a) Time evolution of $\tr\{Q_{S}\rho\}$ and (b) of $\tr\{Q_{T}\rho\}$ as predicted by \eqref{KOM} (solid line) and the traditional as well as the Jones-Hore theory (dashed line).}
\label{fig2}
\end{figure}
We thus consider the most basic case of no magnetic interactions and a single recombination channel, for example the singlet, i.e. we completely eliminate S-T mixing and we take $k_T=0$. We chose as initial state the coherent S-T superposition \cite{note2} $|\psi_{1}\rangle=(|S\rangle+|T\rangle)/\sqrt{2}$ (a coherent superposition is chosen exactly because it is in this maximum coherence regime $p_{\rm coh}=1$ that the new master equation significantly departs from the previous theory). We plot the expectation values of the projectors $Q_S$ and $Q_T$ in Fig.\ref{fig2}. It is evident that the traditional master equation of spin chemistry predicts that half of the radical-ion pairs stay locked in the non-reacting triplet state, whereas the other half produce singlet recombination products. It is as if at $t=0$ we make a global von-Neumann measurement of the entire population of radical-pairs, which are then projected into either the non-reacting triplet and stay there forever, or into the reacting singlet, with probability of 0.5 for both cases. We will explain shortly how this behavior comes about. In contrast, the master equation \eqref{KOM} predicts that 75\% of the molecules will react (Fig.\ref{fig2}). This is because what actually happens is a continuous weak measurement {\it operating within individual radical-ion pair}, resulting to spin decoherence. Before the latter sets in, radical-ion pairs react through the singlet channel, but as time progresses, the initial coherent state is transformed into an incoherent mixture, so the molecules are {\it gradually} locked in the non-reacting triplet state. This process can be visualized with the diagram shown in Fig.\ref{diagram}, describing single-molecule quantum trajectories. 
\begin{figure}
\includegraphics[width=8.5 cm]{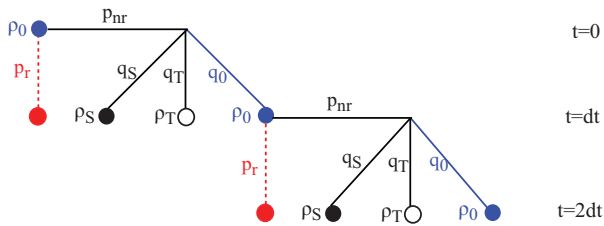}
\caption{(Color online) Single radical-ion-pair state evolution with no singlet-triplet mixing, a single recombination channel ($k_{T}=0$) and a coherent initial state. A radical-ion pair starting out in the state $\rho_{0}$ either recombines (probability $p_r$) or it does not (probability $p_{nr}=1-p_{r}$. If it does not recombine it will either be projected to $\rho_{S}$ (probability $q_S$), or projected to $\rho_{T}$ (probability $q_T$) or else remain at $\rho_{0}$ (probability $1-q_{S}-q_{T}$), and the same scenario unfolds from there on. }
\label{diagram}
\end{figure}
We will trace the state evolution at time steps $dt$, as shown in Fig.\ref{diagram}.
Starting with the coherent state $\rho_{0}=|\psi\rangle\langle\psi|$, with $|\psi\rangle=(|S\rangle+|T\rangle)/\sqrt{2}$, the probability for singlet recombination is $p_{r}=k_{S}dt\langle Q_{S}\rangle=k_{S}dt/2$. In a particular realization of this single molecule experiment this recombination takes place in the first time interval $(0,dt)$, and the reaction terminates. In some other realization the radical-ion pair will not recombine in this first time interval $dt$. Non-recombination happens with probability $p_{nr}=1-p_{r}$. According to \eqref{komME}, spin coherence will be lost for non-recombining radical pairs. In single-molecule realizations, this happens as follows. Conditioned on the fact that the radical-ion pair does not recombine within the interval $(0,dt)$, there are three possibilities for what can happen: (i) a singlet projection to the state $\rho_{S}\equiv Q_{S}\rho_{0}Q_{S}/\tr\{\rho_{0}Q_{S}\}$ takes place with probability $q_{S}=k_{S}dt\tr\{\rho_{0}Q_{S}\}/2=k_{S}dt/4$, (ii) a triplet projection to the state $\rho_{T}\equiv Q_{T}\rho_{0}Q_{T}/\tr\{\rho_{0}Q_{T}\}$ takes place with probability $q_{T}=k_{S}dt\tr\{\rho_{0}Q_{T}\}/2=k_{S}dt/4$, and (iii) the state remains $\rho_{0}$ with probability $q_{0}=1-k_{S}dt/2$ \cite{note}. The average of all three possibilities, for which $q_{S}+q_{T}+q_{0}=1$, reproduces the master equation \eqref{komME} of unrecombined radical pairs. If the radical-pair is projected to the singlet state $\rho_{S}$, it will have $\langle Q_{S}\rangle=1$, so it will recombine for sure at some point. If it is projected to the triplet state $\rho_{T}$, it will remain there forever, as $k_{T}=0$. Finally, if the state remains $\rho_{0}$, the same scenario will unfold in the next time interval. Since those molecules that have been projected to $\rho_{S}$ will definitely recombine at some point, it is easy to compute the singlet yield $Y_{S}$ by summing the total probability for recombining from $\rho_{0}$ with the total probability for being projected to $\rho_{S}$. The probability of the state remaining $\rho_{0}$ is $p_{0}=p_{nr}q_{0}=(1-k_{S}dt/2)^{2}\approx 1-k_{S}dt$, whereas the probability for singlet projection is $p_{S}=p_{nr}q_{S}=q_{S}$. Therefore  
\begin{align}
Y_{S}&=p_{r}+p_{S}+p_{0}(p_{r}+p_{S})+p_{0}^{2}(p_{r}+p_{S})+...\nonumber\\
&={3\over 4}k_{S}dt+(1-k_{S}dt){3\over 4}k_{S}dt+(1-k_{S}dt)^{2}{3\over 4}k_{S}dt+...\nonumber\\
&={3\over 4}
\end{align}
This is how the result shown in Fig.\ref{fig2} is produced, i.e. this is how the total singlet yield comes out to be larger than 0.5, and the fraction of radical-ion pairs remaining locked in the non-reacting triplet state is accordingly smaller than 0.5. At first glance this result sounds unexpected at the least. How can this be acceptable, since we start with 1/2 probability of the radical-pair spin state being singlet or triplet? According to the traditional theory \cite{ivanov}, coherent superpositions of the form $\alpha_{1}|{\rm D}^{\bullet +}{\rm A}^{\bullet -}\rangle+\alpha_{2}|{\rm DA}\rangle$, where $|{\rm D}^{\bullet +}{\rm A}^{\bullet -}\rangle$ is the radical-ion pair state and $|{\rm DA}\rangle$ the neutral product state, are permissible. So the initial state $(|S\rangle+|T\rangle)/\sqrt{2}$ will evolve to $|{\rm singlet~product}\rangle+|{\rm triplet~radical~pair}\rangle)/\sqrt{2}$, hence the expected singlet yield is 0.5, equal to the final fraction of unrecombined triplet molecules. In reality, however \cite{forthcoming}, there cannot be any quantum superpositions between reactants and reaction products of an irreversible exergonic reaction. In other words, there is no physical operator having the radical-ion-pair state and the neutral product state as its eigenvectors. Put differently, there is no physical mechanism performing a global von Neuman measurement that would result into a 50-50 distribution. Hence there is no contradiction with projective measurement quantum mechanics of having a singlet yield larger than 0.5, as is actually the case in this example. In still other words, the intramolecule quantum measurement of $Q_S$ takes place in the radical-pair's spin space and leads to singlet or triplet projections through $Q_{S}$ or $Q_T$. These quantum jumps are completely independent of the recombination events, which can take place at any time, so that the reaction yields cannot be in any physical way considered to represent the result of any kind of measurement \cite{kom_CPL}.
\subsection{Energy and Angular Momentum Conservation}
It might appear that the state of affairs previously outlined might lead to non-conservation of energy or angular momentum. For example, consider adding a singlet-triplet energy splitting produced by a spin-exchange coupling of the form $J\mathbf{s}_{1}\cdot\mathbf{s}_{2}$. Then the energy difference of the singlet and triplet levels would equal $J$, the initial energy of the coherent superposition would be $J/2$ (defining as zero the energy of the singlet level) and it would appear that the final energy balance is broken, since the final triplet population is 25\%, i.e. an energy of $0.25J/2$ per molecule is missing. The same argument would apply for angular momentum conservation, had we considered a superposition of the form $(|S\rangle+|T_{+}\rangle)/\sqrt{2}$, where $T_{+}=|\uparrow\uparrow\rangle$ is the plus-one projection of the total electron spin. We will briefly show why there is actually no violation of these conservation rules. First of all, such superpositions cannot be produced in the first place by any physical magnetic Hamiltonian ${\cal H}$, since the latter will always couple states satisfying energy and angular momentum conservation. Nevertheless, suppose that in some way such superpositions are produced. We introduce a more general definition of $p_{\rm coh}$:
\beq
p_{\rm coh}(t)={{|\langle\langle\tr\{\rho_{ST}(t)\rho_{TS}(t+\tau)\}\rangle\rangle|}\over {\tr\{\rho_{SS}\}\tr\{\rho_{TT}\}}}\label{pcohe}
\eeq
where $\langle\langle..\rangle\rangle$ denotes a time average over $\tau$ (with $\tau$ being larger than the inverse S-T energy difrerence and smaller than characteristic time-scale of the reaction) and $\rho_{TS}(t+\tau)=e^{-i{\cal H}\tau}\rho_{TS}(t)e^{i{\cal H}\tau}$. In the above mentioned case of a singlet-triplet splitting by $J$, the term inside the time-average becomes $e^{-iJt}$, which quickly rotates in the complex plane at an angular frequency $J$, making $p_{\rm coh}\approx 0$. In the $p_{\rm coh}=0$ case we retrieve the traditional theory, which does conserve energy. Essentially, an S-T energy difference makes the two "parties" in the quantum superposition distinguishable, suppressing the effect of quantum interference. Similar considerations apply to angular momentum conservation.
\section{Analogy with the double-slit experiment} 
\begin{figure}
\includegraphics[width=7.5 cm]{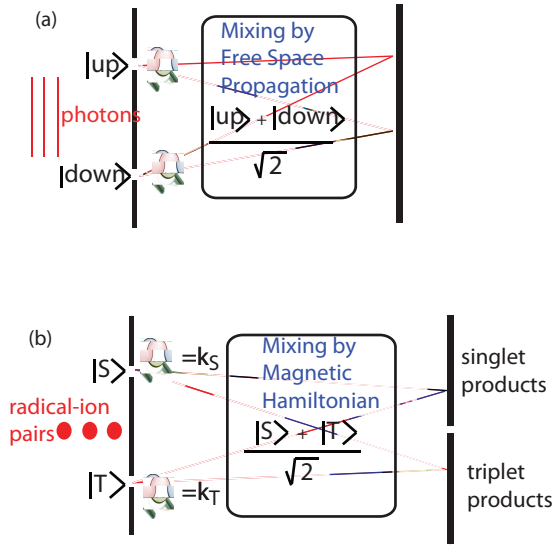}
\caption{(Color online) Analogy between (a) the photon double slit experiment and (b) spin-selective radical-ion-pair reactions. The magnifying glass denotes a detector acquiring "which-path" information. In the radical-ion-pair case this detector is realized by the singlet and triplet reservoir states, which continuously "measure" the pair's spin state at a rate $k_{S}/2$ and $k_{T}/2$, respectively.}
\label{fig3}
\end{figure}

We will now further dwell on several points concerning the analogy with the photon's double slit experiment, schematically depicted in Fig.\ref{fig3}. This analogy will provide a comprehensive understanding of the quantum dynamics of spin-selective radical-ion-pair reactions.
\subsection{Creation of Singlet-Triplet Coherence}
In the optical double slit experiment coherent superpositions are created by free space propagation (i.e. diffraction) across the slits. In radical-ion pairs coherent S-T superpositions are created by the magnetic interactions embodied in the 
magnetic interaction Hamiltonian ${\cal H}$ responsible for the unitary evolution of $\rho$ in \eqref{KOM}. This can be easily shown as follows. We already pointed out that the density matrix can be written as $\rho=\bar{\rho}+\tilde{\rho}$, where
$\bar{\rho}=Q_{S}\rho Q_{S}+Q_{T}\rho Q_{T}$ is the incoherent and $\tilde{\rho}=Q_{S}\rho Q_{T}+Q_{T}\rho Q_{S}$ is the coherent part of $\rho$. The Lindblad term $Q_{S}\rho+\rho Q_{S}-2Q_{S}\rho Q_{S}$ appearing in the master equation \eqref{komME} is equal to $\tilde{\rho}$, i.e. as already explained, the measurement dynamics dissipate singlet-triplet coherence. The unitary part of the master equation \eqref{komME}, $-i[H,\rho]$ is the generator of S-T coherence. Indeed, since $Q_{S}\tilde{\rho}Q_{T}+Q_{T}\tilde{\rho}Q_{S}=\tilde{\rho}$, it follows from \eqref{komME} that 
\beq
{{d\tilde{\rho}}\over {dt}}=-iQ_{S}[{\cal H},\rho]Q_{T}-iQ_{T}[{\cal H},\rho]Q_{S}-{{k_{S}+k_{T}}\over 2}\tilde{\rho}\label{eq9}
\eeq
The first term of the first commutator on the right-hand side of \eqref{eq9} is (apart from $-i$) $Q_{S}{\cal H}\rho Q_{T}$, which after inserting $Q_{S}+Q_{T}=1$ between ${\cal H}$ and $\rho$, and taking into account that $Q_{S}^{2}=Q_{S}$ and $Q_{T}^{2}=Q_{T}$, becomes ${\cal H}_{SS}\rho_{ST}+{\cal H}_{ST}\rho_{TT}$, where ${\cal H}_{\alpha\beta}=Q_{\alpha}{\cal H}Q_{\beta}$ with $\alpha,\beta=S,T$. All other terms in the commutators of \eqref{eq9} are treated similarly, leading to
\begin{align}
{{d\tilde{\rho}}\over {dt}}=-{{k_{S}+k_{T}}\over 2}\tilde{\rho}-&i\Big({\cal H}_{TS}\rho_{SS}-\rho_{SS}{\cal H}_{ST}\nonumber\\
&+{\cal H}_{ST}\rho_{TT}-\rho_{TT}{\cal H}_{TS}\nonumber\\
&+{\cal H}_{SS}\rho_{ST}-\rho_{ST}{\cal H}_{TT}\nonumber\\
&+{\cal H}_{TT}\rho_{TS}-\rho_{TS}{\cal H}_{SS}\Big)
\end{align}
Evidently, even if initially $\tilde{\rho}=0$, i.e. there is no S-T coherence, $\rho_{SS}$ or $\rho_{TT}$ will be different from zero, since $\tr\{\rho_{SS}\}+\tr\{\rho_{TT}\}=\tr\{\rho\}\geq 0$ , so S-T coherence is generated by ${\cal H}_{ST}$ and ${\cal H}_{TS}$. At the same time, it is dissipated at a rate $(k_{S}+k_{T})/2$.

For simplicity in our previous discussions, i.e. for eliminating the unnecessary complications of including spin interactions we just assumed an initial coherent superposition state without specifying how it was created. This is perfectly adequate for the purposes of illustrating the fundamental quantum dynamics of the system and the conceptual differences from the previous physical paradigm of the traditional theory.
\subsection{What is it that interferes?}
In the general case of partial singlet-triplet coherence, a singlet or triplet reaction product can be traced back to two different origins: (i) a singlet or triplet radical pair and (ii) a singlet-triplet coherent radical pair. By detecting a singlet or triplet reaction product, it is physically impossible to differentiate (i) from (ii), and this is embodied in the update rule $d\rho=(1-p_{\rm coh})d\rho_{\rm incoh}+p_{\rm coh}d\rho_{\rm coh}$. This update rule can also be read as $d\rho=d\rho_{\rm incoh}+p_{\rm coh}(d\rho_{\rm coh}-d\rho_{\rm incoh})$, making evident the analogy with the double slit, i.e. the density matrix changes by the first, incoherent term plus the second term which is proportional to the S-T coherence $p_{\rm coh}$. The change $d\rho$ will update the density matrix $\rho_{t}$, which will govern the creation of the reaction products in the following time interval $dt$. 

However, a point where the analogy with the optical double slit is not one-to-one is the fact that in the radical-ion-pair case there is no equivalent of the "photon-field", hence there is no spatially dependent interference pattern. The distribution of photon clicks on the observation screen has its equivalent on just two numbers: the time-integrated reaction yields. They are given by the combined contribution of all possible ways to reach the final singlet or triplet product, as depicted in the example of Fig.\ref{diagram}. 
\subsection{Which-path information and destruction of interference}
How is this interference suppressed? In the double slit, if there is a photon detector providing definite "which-path" information, the interference pattern will be destroyed. Incomplete "which-path" information will reduce the fringe visibility. Whether a photon detector is placed at the upper or the lower slit is indifferent, what matters is the total measurement strength of the photon's path. In other words, it is immaterial whether a strong measurement is done at the upper slit and a weak measurement at the lower slit or vice versa, it is the total measurement strength that provides  "which-path" information. This is so because in the extreme of zero measurement strength at one slit, i.e. when using just one photon detector e.g. at the lower slit, the absence of a detection implies that the photon went through the upper slit, so "which-path" information is still available. In radical-ion pairs, it is the measurement rates $k_{S}$ and $k_{T}$ that gradually provide "which-path" information and reduce $p_{\rm coh}$. Along the reaction, the internal decoherence process (a single-molecule process), taking place at the rate $(k_{S}+k_{T})/2$ will gradually set in and damp singlet-triplet coherence that is generated by the magnetic interactions. In this process it does not matter whether $k_{S}$ is small and $k_{T}$ large or vice versa. What matters is the total measurement rate $(k_{S}+k_{T})/2$ of the observable $Q_{S}$, and this is what enters in Eq. \eqref{komME}. 
\subsection{Interference is a one-particle effect}
Radical-ion pair recombination is a {\it single molecule process}, similar to the double slit quantum interference which {\it is a single photon effect}. In that case, the detection of a photon on the observation screen after the two slits does in no way affect the quantum state of the rest of the photons. However, such a statement is embedded in the traditional theory. Here it is why. Suppose that at time $t$ we have $N$ radical-ion pairs all prepared in the coherent state $|\psi_{1}\rangle=(|S\rangle+|T\rangle)/\sqrt{2}$. Hence at this very instant in time $\rho=N\rho_{1}$, where $\rho_{1}=(|S\rangle\langle S|+|T\rangle\langle T|+|S\rangle\langle T|+|T\rangle\langle S|)/2$ is the single-molecule density matrix. No matter what the master equation satisfied by $\rho$ is, we can obviously write that the change of $\rho$ in the following time interval $dt$ is $d\rho=\rho_{1}dN+Nd\rho_{1}$. The first term represents the change in the density matrix due to reaction and the second the state change of the unrecombined radical pairs. Let us compare the predictions of our approach against the traditional theory. Since $\langle Q_{S}\rangle_{t}=1/2$, both theories predict $dn_{S}=Nk_{S}dt/2$ singlet products during $dt$. We have the machinery to calculate $d\rho_{1}$. Indeed, from \eqref{komME} we find $d\rho_{1}=-(k_{S}dt/2)\rho_{1,{\rm coh}}$, and hence
\beq
d\rho=-dn_{S}\rho_{1}-{{k_{S}dt}\over 2}\rho_{\rm coh}\label{dr_kom}
\eeq
where $\rho_{\rm coh}=N\rho_{1,{\rm coh}}$ and $\rho_{1,{\rm coh}}=(|S\rangle\langle T|+|T\rangle\langle S|)/2$ is the coherence of $\rho_{1}$. Equivalently, we can arrive at \eqref{dr_kom} starting from the general master equation \eqref{KOM} and noting that at this instant $p_{\rm coh}=1$. The interpretation of \eqref{dr_kom} is that we lose $dn_{S}$ radical-ion pairs and the rest (unrecombined) lose some of their coherence. In contrast, the traditional master equation predicts that
\beq
d\rho=-dn_{S}\rho_{1}+N{k_{S}dt\over 4}(|T\rangle\langle T|-|S\rangle\langle S|)\label{drtrad}
\eeq
Here the second term is highly problematic. Identifying this term with $Nd\rho_{1}$, it follows that the unrecombined molecules have become more triplet and less singlet. The analogous statement in the double slit experiment would be that acquiring "which-path" information with a photon detector placed at the upper slit (only $k_{S}\neq 0$) and measuring a particular number of "clicks" on the observation screen, renders the rest of the photons flying through the screen more "down" and less "up". This is impossible. In reality, if "which-path" information is acquired, the photons originally in the state $(|{\rm up}\rangle+|{\rm down}\rangle)/\sqrt{2}$ loose their coherence, i.e. their state approaches the mixed state ${1\over 2}|{\rm up}\rangle\langle{\rm up}|+{1\over 2}|{\rm down}\rangle\langle{\rm down}|$, and that's why the fringe visibility is lost. The equivalent situation in radical-pairs is the loss of spin coherence described in \eqref{dr_kom}. 
\subsection{On the non-linearity of the master equation}
Finally, we comment on the non-linear nature of \eqref{KOM}, which should not come as a surprise \cite{ottinger}. We saw that if the initial state is $|\psi_{1}\rangle=(|S\rangle+|T\rangle)/\sqrt{2}$, 25\% of the molecules are finally locked in the non-reacting triplet state.  For a second ensemble initially in the state $|\psi_{2}\rangle=(|S\rangle-|T\rangle)/\sqrt{2}$, we arrive at the same result. Now imagine mixing these two ensembles. If we naively take for the resulting system $\rho=\rho_{1}+\rho_{2}$, we find $\rho=|S\rangle\langle S|+|T\rangle\langle T|$, i.e. the final ensemble appears to be an incoherent singlet-triplet mixture. Since only the singlet reacts, at the end we would expect to have $\rho_{t=\infty}=|T\rangle\langle T|$, i.e. 50\% of the total number of radical-ion pairs do not react. So what is the final triplet population, 25\% or 50\% ? Let us answer by asking a different question: did the fact that we added the contents of two boxes together destroy the singlet-triplet quantum coherence? Not really. The ensemble $\rho_{\rm prop}=\rho_{1}+\rho_{2}$, specifically prepared as described previously, is a {\it proper mixture} \cite{wiseman,schlossRMP,schlossBOOK}, i.e. each and every molecule is in some coherent S-T superposition, either $|\psi_{1}\rangle$ or $|\psi_{2}\rangle$. We just don't know which. But the reaction does. The reaction is a single-molecule process, {\it not some global process acting on all radical-ion pairs altogether}. So at the end, we will indeed be left with 25\% of the molecules we started with, whether we mix the two ensembles or not. On the other hand, if we start with the two ensembles $|\psi_{1}\rangle$ and $|\psi_{2}\rangle$ and apply a decoherence process, so that the coherent superpositions are turned into incoherent mixtures, we would again obtain $\rho_{\rm improp}=|S\rangle\langle S|+|T\rangle\langle T|$. This time however, we have an improper mixture, not due to state ignorance, but due to a genuine loss of coherence, and we would get 50\% singlet products and 50\% triplet radical pairs. In general, one must clearly define how the initial state of a radical-ion-pair ensemble is prepared in order not to arrive at inconsistencies between theoretical predictions and observations. 

Returning to the double slit analogy, the optical interferometer is linear since the interference pattern is dependent on the phase of the light field. Indeed, suppose we do the experiment in two consecutive runs with equal number of photons in each run. If for the second run we insert a $\pi$ phase shifter after one of the slits, the interference pattern will shift, and adding both runs there will be no interference pattern. That does not mean that the photon coherence was destroyed. This is the analog of the proper mixture of radical-ion pairs, where the singlet-triplet coherence is actually there. However, in this case, the reaction is a non-linear interferometer, so a phase change of the initial state does not alter the reaction yields ($dn_{S}$, $dn_{T}$ and $p_{\rm coh}=1$ are the same for both $|\psi_{1}\rangle$ and $|\psi_{2}\rangle$). If on the other hand, we destroy the coherence $(|{\rm up}\rangle+|{\rm down}\rangle)/\sqrt{2}$ of the photon state, e.g. with a detector providing "which-path" information, the interference pattern will genuinely disappear. This is the analog of the improper mixture of radical-ion pairs, where some decoherence process (for example the unavoidable one internal in the molecules) irreversibly destroys the coherence. 
\section{Conclusions}
Concluding, we have elucidated the fundamental role spin coherence plays in determining the fate of radical-ion-pair reactions, which were shown to be a non-linear spin-coherence interferometer, analysed in direct analogy with the optical double slit experiment. We have derived the fundamental master equation describing these biologically significant biochemical reactions, shown to be the analog of the double slit experiment with partial "which-path" information, stemming from the singlet-triplet decoherence brought about by the intra-molecule quantum measurement dynamics. As long as there are additional spin relaxation mechanisms, that suppress the singlet-triplet coherence $p_{\rm coh}$, or in general, as long as $p_{\rm coh}$ is small, the traditional theory of spin chemistry is expected to be successful. Significant deviations are expected to occur in situations where $p_{\rm coh}$ is appreciable.

\appendix
\section{Properties of $p_{\rm coh}$}
For the maximum incoherence extreme we have $\rho=Q_{S}\rho Q_{S}+Q_{T}\rho Q_{T}$, i.e. by definition $\rho_{ST}=\rho_{TS}=0$, hence $p_{\rm coh}=0$. A maximally coherent state can be written as $\rho=|\psi\rangle\langle \psi|$ with $|\psi\rangle=\alpha |S\rangle+\beta |T\rangle$, and it is easily seen that $\tr\{\rho_{ST}\rho_{TS}\}=|\alpha|^{2}|\beta|^{2}$, whereas $\tr\{\rho_{SS}\}=|\alpha|^{2}$ and $\tr\{\rho_{TT}\}=|\beta|^{2}$, hence indeed $p_{\rm coh}=1$. The general case is treated as follows. Since $\rho_{ST}$ and $\rho_{TS}$ are traceless, $\tr\{\rho\}=\tr\{\rho_{SS}\}+\tr\{\rho_{TT}\}$, thus $\tr\{\rho\}^{2}=\tr\{\rho_{SS}\}^{2}+\tr\{\rho_{TT}\}^{2}+2\tr\{\rho_{SS}\}\tr\{\rho_{TT}\}$. Second, since the projectors $Q_S$ and $Q_T$ are orthogonal, it is readily shown that $\tr\{\rho^{2}\}=\tr\{\rho_{SS}^{2}\}+\tr\{\rho_{TT}^{2}\}+2\tr\{\rho_{ST}\rho_{TS}\}$. But $\tr\{\rho^{2}\}\leq \tr\{\rho\}^{2}$, hence
\beq
{{\tr\{\rho_{SS}^{2}\}+\tr\{\rho_{TT}^{2}\}+2\tr\{\rho_{ST}\rho_{TS}\}}\over {\tr\{\rho_{SS}\}^{2}+\tr\{\rho_{TT}\}^{2}+2\tr\{\rho_{SS}\}\tr\{\rho_{TT}\}}}\leq 1
\eeq
The first two terms in the nominator are less or equal the corresponding first two terms of the denominator, hence the same must hold true for the third terms (otherwise the ratio would become larger than one in the case of a pure state where $\tr\{\rho_{SS}^{2}\}=\tr\{\rho_{SS}\}^{2}$ and $\tr\{\rho_{TT}^{2}\}=\tr\{\rho_{TT}\}^{2}$). We have thus shown that 
\beq
0\leq p_{\rm coh}={{\tr\{\rho_{ST}\rho_{TS}\}}\over {\tr\{\rho_{SS}\}\tr\{\rho_{TT}\}}}\leq 1
\eeq
To motivate the definition of $p_{\rm coh}$, we consider the case of mixing $N$ molecules in the singlet state, described by $\rho=|S\rangle\langle S|$ with another $N$ molecules in the coherent superposition $\rho=|\psi\rangle\langle\psi |=(|S\rangle\langle S|+|T\rangle\langle T|+|S\rangle\langle T|+|T\rangle\langle S|)/2$. Take $k_{S}dt=1/2$, so the probability for the former to recombine within $dt$ is 1/2, while the respective probability for the latter is half as much, i.e. 1/4. If we detect a singlet recombination product, it 
will have thus originated with probability 2/3 from the singlet molecules and with probability 1/3 from the S-T coherent molecules. The latter probability is indeed derived from the expression for $p_{\rm coh}$.


\begin{thebibliography}{0}
\bibitem{note0}
Since the energy of the neutral triplet product (DA)$^{\rm T}$ is in general different from the ground state energy  of DA, these photons or phonons are in principle distinguishable.
\bibitem{note1}
We neglect the presence of nuclear spins and consider only the singlet $|S\rangle$ and the zero projection $|T_{0}\rangle$ state of the triplet manifold, which for notational simplicity is denoted by $|T\rangle$.
\bibitem{note2}
There is nothing special with this particular coherent superposition. It is just chosen as an illustrative example among the infinite number of maximally coherent states of the form $(|S\rangle+e^{i\phi}|T\rangle)/\sqrt{2}$ belonging to the two-dimensional space we are working with.
\bibitem{note}
In the example we are considering only the singlet recombination channel is open ($k_{T}=0$). For completeness we note that in the general case where $k_{S},k_{T}\neq 0$ these probabilities would be $q_{S}=(k_{S}+k_{T})dt\tr\{\rho_{0}Q_{S}\}/2$, $q_{T}=(k_{S}+k_{T})dt\tr\{\rho_{0}Q_{T}\}/2$ and $q_{0}=1-(k_{S}+k_{T})/2$, i.e. as explained in the text, both channels bring about a measurement of the same observable, e.g. $Q_{S}$, at a total measurement rate $(k_{S}+k_{T})/2$.
\bibitem{trad}
All predictions of the traditional theory of spin chemistry are based on the master equation $d\rho/dt=-i[{\cal H},\rho]-{k_{S}\over 2}(Q_{S}\rho+\rho Q_{S})-{k_{T}\over 2}(Q_{T}\rho+\rho Q_{T})$.
\end{thebibliography}
\end{document}